\documentstyle[aps,epsf]{revtex}
\def \be   {\begin{equation}}
\def \ee   {\end{equation}}
\def \l {\label}
\begin{document}
\input epsf
%\hfill {UFES-DF-MMS.01/02}\\
%\medskip
%\hfill \hep-th/yymmxxx
\baselineskip=25pt
%\draft
%\preprint
\title{Gravity and Antigravity with Discrete Interactions: Alternatives I and II}
\author{Manoelito M de Souza\footnote{Permanent address:Departamento de
F\'{\i}sica - Universidade Federal do Esp\'{\i}rito Santo\\29065.900 -Vit\'oria-ES-Brazil- E-mail: manoelit@cce.ufes.br\\Partial grant from the Conselho Nacional de Desenvolvimento Cient\'{\i}fico-CNPq}}
\address{Centro Brasileiro de Pesquisas F\'{\i}sicas- CBPF\\R. Dr. Xavier Sigaud 150,\\
22290-180 Rio de Janeiro -RJ - Brazil}
\date{November 8, 2001}
\maketitle
\begin{abstract}
Questioning the experimental basis of continuous descriptions of fundamental interactions we discuss classical gravity as an effective continuous first-order approximation of a discrete interaction. The sub-dominant contributions produce a residual interaction that may be repulsive and whose physical meaning is of a correction of the excess contained in the continuous approximation. These residual interactions become important (or even dominate) at asymptotical conditions of very large distances from where there are data (rotation curves of galaxies, inflation, accelerated expansion, etc) and cosmological theoretical motivations that suggest new physics (new forms of interactions) or new forms (dark) of matter and energy. We show that a discrete picture of the world (of matter and of its interactions) produce, as an approximation, the standard continuous picture  and more. The flat rotation curve of galaxies, for example, may have a simple and natural explanation.
\end{abstract}
\begin{center}
\end{center}
The idea of continuity pervades all modern physics. Matter and its interactions are described by continuous spacetime functions despite experimental evidences of the quantum nature of the world which seems to be made of pointlike elementary objects like quarks, electrons, photons, gluons, etc, and whose interactions are realized through the exchange of some of these discrete elementary objects. 
Interaction continuity is explicit in the very definition of a classical field and pervades quantum field theory which still makes use of continuous fields. This field continuity is source for infinities. 
A lattice structure for spacetime, so often considered in the literature as a way of escaping from these infinities, is not an appropriate approach as it does not reproduce, even in the limit of zero spacing, all the observed spacetime continuous symmetry. Here the spacetime continuity is taken for granted. Discretization in physics is seen, in general, as a simplification of a richer continuous structure; here we prove the opposite, that the continuous-interaction picture may be a simplified approximation of a much richer discrete structure.

The introduction (or just the recognition) of field discreteness cures infinities and other problems of Classical Electrodynamics \cite{hep-th/9610028} and of General Relativity \cite{gr-qc/9810040} whilst their standard continuous fields are retrieved as effective averaged interactions. These results suggest questioning the experimental basis for a continuous description of the world (of matter and of its interactions). Do they, as usually taken as granted, exclude a discrete description? We want to show here that they do not. We discuss how to experimentally discriminate between  continuous and discrete descriptions of classical gravity, as  a particular example. Discrete interactions reproduce, in a limit of large number of interaction events, the continuous fields but leaves a residue from which some experimental evidences may be detected. 

The discrete field concept \cite{hep-th/9610028,gr-qc/9810040} relies on the assumption of an absolutely discrete world - made of discretely interacting pointlike objects, the discrete fields on a flat continuous spacetime. Although a flat spacetime is not, strictly, an observable \cite{hep-th/0103218}, a curved one would imply on a continuous interaction. Any physical object, if not elementary (in the sense of pointlike, structureless) by itself,  is a discrete set of discrete fields. Discrete fields should not be mistaken with classical particles just for being pointlike; they do superpose (present interference effects) for being fields.  All fundamental interactions are interactions between pairs of discrete elementary fields through the exchange of others discrete fields. Any continuity, except of the spacetime, is just apparent, just a matter of scale.  The worldline of a discretely interacting physical object is continuous but  (piecewise) differentiable only during the free propagation between consecutive interaction events. The concept of force or of acceleration makes sense then only as an approximative limiting continuous description in a range where this approximation is  possible. We must deal, instead, with sudden discrete changes of momentum at the interaction events. Interaction strengths are then parameterized by the time intervals $\Delta t_{i}$ between consecutive interaction events and by the sudden change of momentum $\Delta p_{i}$ there (spin has not been considered yet). The ratio $\frac{\Delta p_{i}}{\Delta t_{i}}$ has no special   meaning. 

The much simpler case of a radial motion with a non-relativistic axially symmetric interaction (where the effective acceleration is inversely proportional to the distance) has been considered in \cite{hep-th/0103218}. Here we will consider a non-relativistic radial motion with a discrete interaction whose effective description is an acceleration field that decreases with the inverse of the squared distance. We will assume large distances, weak fields and small velocities so that relativistic corrections may be neglected. Let us consider the gravitational interaction between two macroscopic objects - a large central mass M and a small one m - the Sun and a probe spacecraft, or a galaxy and a star, just for fixing the idea. They are each made of a large number, $N$ and $N'$ respectively, of elementary components. At a fundamental level the interaction description must be made in terms of interacting pairs of these elementary components. Any description in terms of macroscopic parameters of the objects, like their masses, for example, must be regarded as an approximation. An elementary discrete interaction is an exchange of a quantum (discrete) of interaction. We will refer, below in Eqs. (\ref{dv1}) and (\ref{dti1}), to two alternative ansats about discrete gravity but for the second one we will present only the final results  as it is being discussed elsewhere \cite{gr-qc/0111031}. It is included here for the sake of completeness and comparison as both alternatives reproduce the standard continuous fields and both present discrepancies on the same  limiting situations. The point here, it is necessary to make it explicit, is not that gravity be actually described by any of these two alternatives. Both are gross simplifications and have their own limitations. The question is how far experiences back the continuous field concept and exclude discrete alternatives? 

The emitted quantum is the consequence (or the cause) of a sudden change in the state of movement of its emitter (or, respectively, of its absorber):
\be
\l{dv1}
\Delta p_{i}\equiv q,\qquad{\hbox{or alternatively}}\qquad\Delta p_{i}\equiv \frac{q}{r_{i}},
\ee
where $q$ is a constant, $i$ labels the interaction events, and $r_{i}$ is the distance between the two interacting elementary objects. There is an important ingredient of classical causality here: The interaction between two elementary objects is made through the exchange of a quantum of  interaction (discrete field) whose emission is triggered by  the  absorption of a quantum. As the emitted quantum has to propagate from its emitter to reach its absorber there is, consequently, a time interval $\Delta t_{i}$ between two consecutive interactions which, in order to reproduce the observed effective Newtonian acceleration, must then be given by
\be
\l{dti1}
\Delta t_{i}=\frac{\alpha}{NN'} r^{2}_{i},\qquad{\hbox{or alt.}}\qquad\Delta t_{i}=\frac{\alpha}{NN'} r_{i},
\ee
where $\alpha$ is 
another constant. Eq. (\ref{dti1}) implies on two largely distinct time-scales of interactions: the one for interactions between neighbouring elements of a same object is infinitely smaller than the one for interactions between elements of distinct macroscopic objects. This must be considered when discussing just one of them. When the spatial extensions of the interacting macroscopical objects are much smaller than their space separation $r_{i}$, we can , as usual, consider in an approximation all the components of each of them at a same position in space but they cannot be replaced by single mass-equivalent components having their total masses. The numbers of components, in Eq. (\ref{dti1}), are essential for the interaction description but the fact that $N$ and $N'$ are not an available information eliminates a subtlety that gravity might be determined by the numbers of elementary components and not by their mass-energy.

The presence of $NN'$ in Eq. (\ref{dti1}) reveals the statistical-average-time character of $\Delta t_{i}$. $N$ is equal to unity for an electron, for example, but not for a nucleon as this is structured in terms of point objects - three quarks and an unknown number of gluons. It is impossible to have $N$ and $N'$ as exact known numbers but we can work with an average mass $\mu$ for the elementary constituents
\be
\mu:=\frac{M}{N'}:=\frac{m}{N},
\ee
with the assumption that all the interacting objects are made of the same fundamental stuff and roughly in the same proportion.  The universality of $\mu$ corresponds to the validity of the Equivalence Principle which has an experimental confirmation at the level of precision of $10^{-12}$ \cite{Y}. Therefore, in both cases
\be
\l{LN}
\frac{\Delta p_{i}}{\Delta t_{i}}=\frac{qNN'}{\alpha}\frac{1}{r_{i}^2}=\frac{q}{\alpha\mu^2}\frac{Mm}{r_{i}^2},\quad G:=\frac{q}{\alpha\mu^2},
\ee
but we can only speak of the LHS as a force in an approximated continuous limit \cite{hep-th/0103218} when this ratio may effectively be replaced by a derivative which requires that both $\Delta p_{i}$ and $\Delta t_{i}$ tends together to zero in a such way that their ratio exist. The key point here is that the spacetime is continuous and so, whereas $r_{i}$ can smoothly go to zero $\Delta p_{i}$ cannot because of its intrinsical discrete character. So the derivative fails as a valid approximation in two significative situations and in  both  ``new physics" appears as a departure from the standard continuous picture. The first one is for $r_{i}$ so close to zero that $\Delta p_{i},$ (either $q$ or $\frac{q}{r_{i}}$)  cannot w.r.t. $\Delta t_{i}$ be regarded as an infinitesimal increment; then the departure from the finite discrete interaction appears as false infinities or false new-short-distance  fundamental interactions (fifth forces) in the effective continuous descriptions. This could be taken as an easy basis for an experimental decision on the nature of the fundamental physical interactions but the concept of renormalization acts as a rescue for the continuous interaction hypothesis. So, the question passes to be if the need of renormalization in theories of fundamental interactions is required by Nature (by its here questioned continuous character) or just by the continuous way used to describe it. The other case is for $r_{i}$ so large that  $\Delta t_{i}$ cannot relatively be considered a small increment; it may even be indirectly detectable.  We are interested here on the tiny residual interactions from huge masses (stars and galaxies) through very large distances. Such  macroscopic objects exclude any concern about quantum mechanics and quantum fluctuations. A classical approach is justified.

For the probe initial conditions taken as $r(t_{0})= r_{0}$ and $p(t_{o})=p_{0},$ the next interaction will occur at 
$t_{1}= t_{0}+\Delta t_{0},$
with 
$p(t_{1})\equiv p_{1}=p_{0}-\Delta p_{1}$,
and
$r(t_{1})\equiv r_{1}= r_{0}+\frac{p_{0}}{m}\Delta t_{0},
$
as we are neglecting relativistic corrections, and there is free propagation between  consecutive interaction events.  At the $n^{th}$ interaction, according to the first alternatives in Eqs (\ref{dv1}) and (\ref{dti1}),
\be
\l{vn}
p_{n}=p_{0}-\sum_{i=1}^{n}\Delta p_{i}=p_{0}-nq,
\ee
$q>0$, or $q<0$, in the assumption that the probe is moving, respectively, away from or towards the central mass.
\be
\l{rn}
r_{n}=r_{n-1}+\frac{ p_{n-1}}{m}\Delta t_{n-1}=r_{n-1}+\frac{\alpha }{mNN'}r^2_{n-1}p_{n-1},
\ee
which, recursively, produces 
\be
\l{xn0}
x_{n}=x_{n-1}(1+\beta x_{n-1}p_{n-1})=\prod_{j=0}^{n-1}(1+\beta p_{j}x_{j})
\ee
with $x_{n}:=\frac{r_{n}}{r_{0}}$ and $\beta:=\frac{r_{0}\alpha}{mNN'}=\frac{r_{0}\alpha\mu^2}{Mm^2}=\frac{r_{0}q}{GMm^2}$. The Eqs. (\ref{dv1})  and (\ref{dti1}) replace the differential equations of the continuous fields and the finite series of Eqs. (\ref{vn}) and (\ref{rnA}) replace their respective continuous solutions.
 Expanding the RHS of Eq. (\ref{xn0}) we are led to
\be
\l{T}
x_{n}=1+\beta\sum^{n-1}_{j_{1}=0}p_{j_{1}}x_{j_{1}}+\beta^2\sum^{n-2}_{j_{1}=0}\sum^{n-1}_{j_{2}=j_{1}+1}p_{j_{1}}x_{j_{1}}p_{j_{2}}x_{j_{2}}+\dots+\beta^n\sum^{0}_{j_{1}=0}\sum^{1}_{j_{2}=j_{1}+1}\dots\sum^{n-1}_{j_{n}=j_{n-1}+1} p_{j_{1}}x_{j_{1}}\dots p_{j_{n}}x_{j_{n}}.
\ee
We should observe in this equation, before proving it, that the inferior limit of each sum in the products of sums is given by the value of the variable in the precedent sum added of one unit, starting with zero in the first sum; as a consequence the superior limit of each sum is determined by the superior limit of the subsequent sum diminished of one unit, starting with $n-1$ in the last sum. In a product of sums the superior and inferior limits of each sum are determined by the respective limits of the last and of the first sums. From now on, for the sake of a lighter notation, we will write just, the first inferior and the last superior limits of sums in each product.

Assuming the validity of Eq. (\ref{T}) for $x_{n}$ and using the middle term of Eq. (\ref{xn0}) in order to get  the equivalent series for $x_{n+1},$ the coefficient of its generic $\beta^i$-term is
\be
\l{dT}
\sum_{j_{1}=0}\sum_{j_{2}}\dots\sum^{n-1}_{j_{i}} p_{j_{1}}x_{j_{1}}\dots p_{j_{i}}x_{j_{i}}+p_{n}x_{n}\sum_{j_{1}=0}\sum_{j_{2}}\dots\sum^{n-1}_{j_{i-1}} p_{j_{1}}x_{j_{1}}\dots p_{j_{i-1}}x_{j_{i-1}}
=\sum_{j_{1}=0}\sum_{j_{2}}\dots\sum^{n}_{j_{i}} p_{j_{1}}x_{j_{1}}\dots p_{j_{i}}x_{j_{i}},
\ee
which proves Eq. (\ref{T}), by induction.
The recursive use of Eq. (\ref{T}) leads to
\be
\l{xs}
x_{n}=\sum_{s=0}\beta^{s}x_{n}^{(s)}
\ee 
where $x_{n}^{(s)}$ is a polynomial function of $p_{j}$'s. In particular, $x_{n}^{(0)}=1$, and  
\be
\l{xn1ap}
x_{n}^{(1)}=\sum_{j=0}^{n-1}p_{j}=\sum_{j=0}^{n-1}(p_{0}-jq)={n\choose 1}p_{0}-{n\choose 2}q=[n(p_{0}-\frac{nq}{2})]+[\frac{nq}{2})]=[(\frac{p_{0}^2-p_{n}^2}{2q})]+[(\frac{nq}{2})],
\ee
where the brackets separate the dominant from the sub-dominant contributions from the combinatorials.

It is a simple matter to show that the superior limit of the sum in Eq. (\ref{xs}) is given by $\sum^{n}_{j=1}{n\choose j}$, but since this information is of no further use here it will be just omitted. Using the Eq. (\ref{xs}) to replace each $x_{j}$ in Eq. (\ref{T}) we see that each term of this series (\ref{T}) gives a specific contribution to $x_{n}^{(s)}.$ If we define
\be \l{xdx}
(x_{j_{1}}x_{j_{2}}\dots x_{j_{k-1}}x_{j_{k}})^{(i)}:=\sum_{m=0}^{i}(x_{j_{1}}x_{j_{2}}\dots x_{j_{k-1}})^{(i-m)}x_{j_{k}}^{(m)},
\ee
the contribution from the $\beta^{i}$-term of the series (\ref{T}) to $x_{n}^{(s)}$ is 
\be
\sum_{j_{1}=0}\sum_{j_{2}}\dots \sum^{n-1}_{j_{i}}
p_{j_{1}}p_{j_{2}}\dots p_{j_{i}}(x_{j_{1}}x_{j_{2}}\dots x_{j_{i}} )^{(s-i)}\quad\hbox{for}\quad i\le s.
\ee
Then
we have 
\be
\l{xns}
x_{n}^{(s)}=\sum^{n-1}_{j_{1}=0}p_{j_{1}}x_{j_{1}}^{(s-1)}+\sum_{j_{1}=0}\sum^{n-1}_{j_{2}}p_{j_{1}}p_{j_{2}}(x_{j_{1}}x_{j_{2}})^{(s-2)}+\dots+\sum_{j_{1}=0}\sum_{j_{2}}\dots \sum^{n-1}_{j_{s}}
p_{j_{1}}p_{j_{2}}\dots p_{j_{s}}(x_{j_{1}}x_{j_{2}}\dots x_{j_{s}} )^{(0)},
\ee
for $s>0$,
with the convention that $x_{n}^{(s)}=0$ for $s<0$.

We have dealt, up to here, with exact and rigourous expressions but in all cases of physical interest $n$ is a very huge number and so, for $s<<n$, the replacement in Eq. (\ref{xns}) of the superior limit of each sum, i.e. $n-1,\; n-2,\dots,n-s$,  just by $n$ is a very good approximation. We will let it explicit replacing $x_{n}^{(s)}$ with this approximation by $y_{n}^{(s)}$; it allows the use of a more compact notation for Eq. (\ref{xns}):
\be
\l{T1}
y_{n}^{(s)}=\sum^{n}_{j_{1}=0}p_{j_{1}}{\Big\{}y_{j_{1}}^{(s-1)}+\sum^{n}_{j_{2}}p_{j_{2}}{\Big\{}(y_{j_{1}}y_{j_{2}})^{(s-2)}+\sum^{n}_{j_{3}}p_{j_{3}}{\Big\{}\dots+\sum^{n}_{j_{s-1}}
p_{j_{s-1}}(y_{j_{1}}y_{j_{2}}\dots y_{j_{s-1}} )^{(0)}{\Big\}}\dots{\Big\}}\quad\hbox{for}\quad s<<n
\ee
We observe then the following useful property 
\be
\l{xn1}
y_{n}^{(1)}=\sum_{j_{1}=0}^{n}p_{j_{1}}=(\sum_{j_{1}=0}^{j}+\sum_{j_{1}=j+1}^{n})p_{j_{1}}= y_{j}^{(1)}+\sum_{j_{1}=j+1}^{n}p_{j_{1}}\quad\hbox{for any}\quad j<n.
\ee
Let us see now the structure of $y_{n}^{(2)}$
\be
y_{n}^{(2)}=\sum^{n}_{j_{1}=0}p_{j_{1}}{\Big\{}y_{j_{1}}^{(1)}+\sum^{n}_{j_{2}}p_{j_{2}}{\Big\}}=y_{n}^{(1)}\sum^{n}_{j_{1}=0}p_{j_{1}}= (y_{n}^{(1)})^{2},
\ee
where we made use of Eqs. (\ref{xn1}). Let us assume that
\be
\l{xnsg}
y_{n}^{(k)}= (y_{n}^{(1)})^{k}\quad\hbox{for}\;k=0,1,2\dots ,s\quad \hbox{and}\;s<<n.
\ee
Then we will prove it for $y_{n}^{(s+1)}$ using  Eqs. (\ref{T1}) and (\ref{xn1}) for writing 
$$
\l{nsn1}
y_{n}^{(s)}y_{n}^{(1)}=\sum_{j_{1}=0}^{n}p_{j_{1}}{\Big\{} y_{j_{1}}^{(s-1)}(y_{j_{1}}^{(1)}+\sum^{n}_{j_{2}}p_{j_{2}})+ \sum^{n}_{j_{2}}p_{j_{2}}{\Big\{}(y_{j_{1}}y_{j_{2}})^{(s-2)}(y_{j_{2}}^{(1)}+\sum^{n}_{j_{3}}p_{j_{3}})+\sum^{n}_{j_{3}}p_{j_{2}}{\Big\{}(y_{j_{1}}y_{j_{2}}y_{j_{3}})^{(s-3)}(y_{j_{3}}^{(1)}+\sum^{n}_{j_{4}}p_{j_{4}})+\dots$$
\be
\l{T2}
\dots+\sum^{n}_{j_{s-1}}
p_{j_{s-1}}{\Big\{}(y_{j_{1}}y_{j_{2}}\dots y_{j_{s-1}} )^{(1)}(y_{j_{s-1}}^{(1)}+\sum^{n}_{j_{s}}p_{j_{s}})+\sum^{n}_{j_{s}}
p_{j_{s}}(y_{j_{s}}^{(1)}+\sum^{n}_{j_{s+1}}p_{j_{s+1}}){\Big\}}\dots{\Big\}}.
\ee
From the definition (\ref{xdx}),
\be
(y_{j_{1}}\dots y_{j_{i+1}})^{(s-i-1)}y_{j_{i+1}}^{(1)}=\sum^{s-i-1}_{m=0}(y_{j_{1}}\dots y_{j_{i}})^{(s-i-m-1)}y^{(m+1)}_{j_{i+1}}=\sum^{s-i}_{m=1}(y_{j_{1}}\dots y_{j_{i}})^{(s-i-m)}y^{(m)}_{j_{i+1}},
\ee
and then
\be
\l{gi}
(y_{j_{1}}\dots y_{j_{i}})^{(s-i)}+(y_{j_{1}}\dots y_{j_{i+1}})^{(s-i-1)}y_{j_{i+1}}^{(1)}=\sum^{s-i}_{m=0}(y_{j_{1}}\dots y_{j_{i}})^{(s-i-m)}y^{(m)}_{j_{i+1}}=(y_{j_{1}}\dots y_{j_{i+1}})^{(s-i)}.
\ee
Using Eq. (\ref{gi}) in Eq. (\ref{T2}) we have
$$
(y_{n}^{(1)})^{s+1}=\sum^n_{j_{1}}p_{j_{1}}{\Big\{} y_{j_{1}}^{(s)}+ \sum^{n}_{j_{2}}p_{j_{2}}{\Big\{}(y_{j_{1}}y_{j_{2}})^{(s-1)}+\sum^{n}_{j_{3}}p_{j_{2}}{\Big\{}(y_{j_{1}}y_{j_{2}}y_{j_{3}})^{(s-2)}+\dots$$
\be
\l{T3}
\dots+\sum^{n}_{j_{s}}
p_{j_{s}}{\Big\{}(y_{j_{1}}y_{j_{2}}\dots y_{j_{s}} )^{(1)}+\sum^{n}_{j_{s}}
p_{j_{s}}{\Big\}}\dots{\Big\}}:=y_{n}^{(s+1)},
\ee
where we have used Eq. (\ref{T1}) at the end for the definition of $y_{n}^{(s+1)}$. This proves
\be
\l{y}
y_{n}^{(s+1)}=(y_{n}^{(1)})^{s+1},\quad\hbox{for all integers $s$,}\;s<<n.
\ee
Let us keep track of the smaller-order terms that are being neglected in $y_{n}^{(s)}$ from $x_{n}^{(s)}$ denoting them by $\delta x_{n}^{(s)}$, 
\be
\delta x_{n}^{(s)}\equiv x_{n}^{(s)}-y_{n}^{(s)},
\ee
We have, for example,
\be
\l{dxn1}
\delta x_{n}^{(1)}=x_{n}^{(1)}-y_{n}^{(1)}=(\sum_{j=0}^{n-1}-\sum_{j=0}^{n})p_{j}=-p_{n},
\ee
\be
\delta x_{n}^{(2)}=x_{n}^{(2)}-y_{n}^{(2)}=-2x_{n}^{(1)}p_{n}-\sum_{j=0}^{n}p^2_{j}.
\ee
Then, from Eqs. (\ref{xn1ap}) and (\ref{y}),
\be
\l{lims}
x_{n}=\sum_{s=0}\beta^{s}(y_{n}^{(s)}+\delta x_{n}^{(s)})=\sum_{s=0}\beta^{s}((y_{n})^{s}+\delta x_{n}^{(s)})\approx\frac{1}{1-\beta y_{n}^{(1)}}+\sum_{s=0}\beta^{s}\delta x_{n}^{(s)},
\ee
where, in the last step, a finite series (with $\sum_{j=0}^{n}{n\choose j}$ terms) was replaced by the  limiting sum of an asymptotical series (as if $n$ were an infinite number). This approximation is grounded on $n$ being a very large number and on $\beta y_{n}^{(1)}<<1$ as implicitly assumed.
Considering the Eqs. (\ref{xn1ap}) and (\ref{dxn1}), we have 
\be
\l{xn}
\frac{r_{n}}{r_{0}}\approx\frac{1}{1-\beta[\frac{(p_{0}^2-p_{n}^2)}{2q}+(p_{n}+\frac{nq}{2})]}-\beta p_{n}+{\cal O}(\beta^2).
\ee
Keeping it to this order of approximation leads to
\be
r_{n}\approx\frac{r_{0}}{1-\frac{\beta}{q}(\frac{p_{0}^2-p_{n}^2}{2})}+\delta r_{n}, \qquad\delta r_{n}\approx\frac{q}{GMm^2}[r_{n}^2(p_{n}+\frac{nq}{2})-r_{0}^2p_{n}],\qquad{\hbox{or alt.}}\qquad\delta r_{n}\approx\frac{nq^2}{2GMm^2},
\ee
and to
\be
\l{energy}
\frac{p_{n}^2}{2m}-\frac{GMm}{r_{n}}+m\delta U(n,r_{n})=const,
\ee
with
\be
\delta U(n,r_{n})=\frac{q}{m^2}(\frac{r_{0}p_{n}}{r_{n}}+\frac{nq}{2}),\quad{\hbox{or alt.}}\quad\delta U(n,r_{n})=\frac{n}{r_{n}^2}\frac{q^2}{2m}
\ee
which retrieves from the dominant contribution the Newtonian potential $U(r)=-\frac{GM}{r}$ as an effective continuous field, and a new, relatively negligible extra generalized potential $\delta U(r_{n},p_{n})$ from the sub-dominant contributions. The corresponding expressions from the alternative assumptions in Eqs. (\ref{dv1}) and (\ref{dti1}) are being
 included for comparison sake. Energy is always conserved but the continuous function $U(r)$ is valid as the expression of a potential energy only as an asymptotic idealized limit of $n\rightarrow\infty$.

This is not to be seen as an isolated example but as a general rule for all continuity in physics  with the only exception of the spacetime. In an absolutely discrete world (made of discrete objects with discrete interactions) there is no smooth curve but only straight-line segments so that any mathematical relationship expressing a physical phenomenon must be, in its exact form, expressed in terms of a power series of these straight line segments. A continuous field as any non-power series relationship is just an idealized asymptotic form reachable only after infinite interactions. The interested reader is referred to the discussion in Section VI of \cite{hep-th/0103218} where Boltzmann statistics is seen as the asymptotic limit of a power-function based statistics like the Tsallis \cite{Tsallis} one, as an example. The point in this comment is to say that behind the crude  phenomenological hypothesis (\ref{dv1}) and (\ref{dti1}) there is a consistent and rigourously proved field-theoretic structure showing how the continuous field formalism is generated as effective averages from the discrete field formalism \cite{hep-th/9610028}. The Lagrangian formalism with all its implications remains a valid tool; the fields are discrete with respect to their spacetime extension but remains continuous spacetime functions as they are propagating fields. The discrimination between  the continuous and the discrete interaction-descriptions must experimentally be decided by the detection or not of effects from the neglected smaller contributions. As $n$, and therefore $r_{n}$, becomes progressively larger these neglected contributions, which grow also with $n$, may become detectable as a departure from a continuous description. 

For $n>>\Delta n=1$ the use of derivative becomes a justified approximation. So, from Eqs. (\ref{vn}), (\ref{lims}) and (\ref{xn1ap}) we have
\be
\l{dndr}
\frac{dr_{n}}{dn}\approx r_{0}\frac{d}{dn}(\frac{1}{1-\beta y_{n}^{(1)}})\approx\frac{qp_{n}}{|a_{n}|m^2},\qquad{\hbox{or alt.}}\qquad \frac{dr_{n}}{dn}\approx \frac{qp_{n}r_{n}}{GMm^2},
\ee
\be
 \frac{dp_{n}}{dn}\approx-q,\qquad{\hbox{or alt.}}\qquad \frac{dp_{n}}{dn}\approx-\frac{q}{r_{n}},
\ee
and
\be
\frac{d\delta r_{n}}{dn}=\frac{q^2}{m^2|a_{n}|}(-\frac{1}{2}+\frac{r_{0}^2}{r_{n}^2}+\frac{r_{n}p_{n}(2p_{n}+nq)}{GMm^2})\qquad{\hbox{or alt.}}\qquad\frac{d\delta r_{n}}{dn}\approx \frac{q}{2mr_{n}},
\ee
where $a_{n}=-GM/r^2_{n}$ is the Newtonian acceleration at $r_{n}$. If we want to express it in terms of time-derivative, we return to Eq. (\ref{dti1}) from which we get
\be
t_{n}=t_{0}+\frac{q}{GMm}\sum_{i=0}^{n-1}r_{i}^{2},\qquad{\hbox{or alternatively}}\qquad t_{n}=t_{0}+\frac{q}{GMm}\sum_{i=0}^{n-1}r_{i},
\ee
which, for $n>>1$, leads to
\be
dt_{n}\approx\frac{q}{GMm}r_{n}^{2}dn=\frac{q}{m|a_{n}|}dn,\qquad{\hbox{or alternatively}}\qquad dt_{n}\approx\frac{qr_{n}}{GMm}dn .
\ee
Then, from Eq. (\ref{dndr}), for both alternatives
\be
\frac{dr_{n}}{dt_{n}}\approx\frac{p_{n}}{m}\qquad{\hbox{and}}\qquad\frac{d^{\;2}r_{n}}{dt^2_{n}}\approx\frac{1}{m}\frac{dp_{n}}{dn}\frac{dn}{dt_{n}}=a_{n},
\ee
as expected, and
\be
\frac{d\delta r_{n}}{dt_{n}}\approx\frac{q}{m}(-\frac{1}{2}+\frac{r_{0}^2}{r_{n}^2}+\frac{r_{n}p_{n}(2p_{n}+nq)}{GMm^2}),\qquad{\hbox{or alt.}}\qquad\frac{d\delta r_{n}}{dt_{n}}\approx\frac{q}{2mr_{n}},
\ee
and
\be
\frac{d^{\;2}\delta r_{n}}{dt^2_{n}}\approx\frac{2q(p_{0}+p_{n})}{GMm^3}(\frac{p_{n}^2}{2m}-\frac{GMm}{r_{n}})+\frac{q}{m^2r_{n}}(\frac{p_{0}}{r_{n}}-\frac{2p_{n}r_{0}^2}{r_{n}^3}),\quad{\hbox{or alt.}}\quad\frac{d^{\;2}\delta r_{n}}{dt^2_{n}}\approx-\frac{qp_{n}}{2r^2_{n}m^2},
\ee
which shows that the continuous description is entirely generated by the discrete dominant contributions and that the sub-dominant ones generate new physics that may or may not be observed in nature. 
The extra interaction, in a first order approximation gives a positive increment 
$\delta r_{n}$ to the expected (from the continuous approximation) value of $r_{n}$. Whereas it is always attractive for the second alternative, for the first one, depending on the initial conditions ($p_{0}$ and $r_{0}$), the extra interaction may be repulsive: the actual gravitational attraction is weaker (alternatively, stronger) than its continuous description. The extra interaction has this clear physical meaning. It is a correction to the excess contained in approximating the interaction by the effective continuous potential. Only precise, carefully done experiments must give the final answer. In contradiction to all other fundamental interactions the great riddles associated to gravity are in the large distance limit where the field is extremely weak. They are not easily fitted with a continuous interaction without evoking new forms of interaction or of matter. In a natural way discrete interactions introduce modifications to the standard physics in the very large and in the very small distance limits, and it is only in these asymptotic limits that these modifications can be experimentally detected. Outside them $n$ is huge enough for validating the continuous approximation but not enough for turning detectable the subdominant terms ($\delta x_{n}^{(s)}$). A final heuristic comment for a brief indication of how discrete interaction can drastically change a physical picture let us remark that the flat rotation curves of galaxies may have a simpler explanation if the extra potential $\delta U(r_{n},p_{n})$ is just added to the standard potential $U(r_{n})$. Then, the requirement of $\frac{dV}{dr}=0$ the circular-orbit condition leads to a finite $r_{lim}$, instead of the usual infinite value. It is a hint.

\newpage
\begin{center}
Gravity and antigravity with discrete interactions: Alternative II\\
Manoelito M de Souza\footnote{Permanent address:Departamento de
F\'{\i}sica - Universidade Federal do Esp\'{\i}rito Santo\\29065.900 -Vit\'oria-ES-Brazil- E-mail: manoelit@cce.ufes.br\\Partial grant from the Conselho Nacional de Desenvolvimento Cient\'{\i}fico-CNPq}\\
\address{Centro Brasileiro de Pesquisas F\'{\i}sicas- CBPF\\R. Dr. Xavier Sigaud 150,\\
22290-180 Rio de Janeiro -RJ - Brazil}
\end{center}
\date{November 8, 2001}
%\maketitle
\begin{abstract} 
Questioning the experimental basis of continuous descriptions of fundamental interactions we discuss classical gravity as an effective continuous first-order approximation of a discrete interaction. The sub-dominant contributions produce a residual interaction whose physical meaning is of a correction of the excess contained in the continuous approximation. These residual interactions become important (or even dominate) at asymptotical conditions of very large distances from where there are data (rotation curves of galaxies, inflation, accelerated expansion, etc) and cosmological theoretical motivations that suggest new physics (new forms of interactions) or new forms (dark) of matter and energy. We show that a discrete picture of the world (of matter and of its interactions) produce, as an approximation, the standard continuous picture  and more. Here we discuss a second alternative where the time interval between two consecutive discrete interactions is the flight time of the exchanged interaction quantum. 
\end{abstract}
\begin{center}
\end{center}
Discretization in physics is seen, in general, as a simplification of a richer continuous structure; here we prove the opposite, that the continuous-interaction picture may be a simplified approximation of a much richer discrete structure.
We discuss with more calculation details  the alternative II, already discussed in \cite{gr-qc/0111030}. 
We consider a non-relativistic radial motion with a discrete interaction whose effective description is an acceleration field that decreases with the inverse of the squared distance. 
Assuming that the time interval between two consecutive interactions is the (statistical average) two-way flight time between two macroscopic objects with, respectively $N$ and $N'$ elementary point objects (components), then 
\be
\l{dti}
\Delta t_{i}=\frac{\alpha}{NN'} r_{i},
\ee
where $r_{i}$ is their space separation at the $i^{th}$ interaction. This requires that the change in momentum at each interaction event be given by 
\be
\l{dv}
\Delta p_{i}\equiv\frac{q}{r_{i}},
\ee
in order to reproduce the observed effective Newtonian acceleration. See the reference \cite{hep-th/9610145} for a physical interpretation.
$\alpha$ and $q$ are constants to be determined later. The assumption of large distances, weak fields and low velocities justify a non-relativistic approach. The rest frame of a central source, a mass $M$ in the case of gravity, is then assumed. Clearly this is an effective description which is denounced by the singularity (infinity) at the origin. It does not fit the discrete-field philosophy which implies finite interactions but, like the alternative I, both discussed in \cite{gr-qc/0111030}, it reproduces the standard continuous gravity and both present discrepancies on the same  limiting situations.  

The presence of $NN'$ in Eq. (\ref{dti}) reveals the statistical-average-time character of $\Delta t_{i}$. $N$ is equal to unity for an electron, for example, but not for a nucleon as this is structured in terms of point objects - three quarks and an undetermined number of gluons. It is impossible to have $N$ and $N'$ as exact known numbers but we can work with an average mass $\mu$ for the elementary constituents
\be
\mu:=\frac{M}{N'}:=\frac{m}{N},
\ee
with the assumption that all the interacting objects are made of the same fundamental stuffs and roughly in the same proportion.  The universality of $\mu$ corresponds to the validity of the Equivalence Principle which has an experimental confirmation\cite{Y} at the level of precision of $10^{-12}$. Therefore, in both cases
\be
\l{LNA}
\frac{\Delta p_{i}}{\Delta t_{i}}=\frac{qNN'}{\alpha}\frac{1}{r_{i}^2}=\frac{q}{\alpha\mu^2}\frac{Mm}{r_{i}^2},\quad G:=\frac{q}{\alpha\mu^2}
\ee
For initial conditions taken as $$r(t_{0})= r_{0};\qquad p(t_{0})=p_{0},$$ the next interaction will occur at 
\be
\l{Dt0}
t_{1}= t_{0}+\Delta t_{0}\equiv t_{0}+\frac{\alpha}{NN'} r_{0},
\ee
neglecting relativistic corrections.   
 Then, 
\be
p(t_{1})\equiv p_{1}=p_{0}-\frac{q}{r_{1}},
\ee
and
\be
\l{r1}
r(t_{1})\equiv r_{1}= r_{0}+p_{0}\Delta t_{0}=(1+\beta p_{0})r_{0},
\ee
as there is free propagation between any two consecutive interaction events; $\beta=\frac{\alpha}{NN'm}=\frac{q}{GMm^2}$.
Therefore, in the $n^{th}$ interaction 
\be
\l{rnA}
r_{n}=r_{n-1}+p_{n-1}\Delta t_{n-1}=(1+\beta p_{n-1})r_{n-1}=r_{0}\prod_{i=0}^{n-1}(1+\beta p_{i}),
\ee
 with 
\be
\l{vj}
p_{j}=p_{0}-\sum_{i=1}^{j}\frac{q}{r_{i}}.
\ee
Then
\be
x_{n}\equiv\frac{r_{n}}{r_{0}}= 1+\beta\sum_{i_{1}=0}^{n-1}p_{i_{1}}+\beta^2\sum_{i_{1}=0}^{n-2}\sum_{i_{2}=i_{1}+1}^{n-1}p_{i_{1}}p_{i_{2}}\dots+\beta^{n-1}(\sum_{i_{1}=0}^{0}\sum_{i_{2}=i_{1}+1}^{1}\dots \sum_{i_{n-1}=i_{n-2}+1}^{n-1})p_{i_{1}}p_{i_{2}}\dots p_{n-1},
\ee
or, for large $n\;(n>>1)$
\be
\l{rn0}
\frac{r_{n}}{r_{0}}= 1+\beta\sum_{i_{1}=0}^{n}p_{i_{1}}{\Big\{}1+\beta\sum_{i_{2}=i_{1}+1}^{n}p_{i_{2}}{\Big\{}+\dots+\sum_{i_{n-1}=i_{n-2}+1}^{n}p_{n-1}{\Big\}}\dots{\Big\}},
\ee
in a more compact notation.
This is a finite series of $n+1$ terms. Our objective is, of course, writing $r_{n}$ and $p_{n}$ in terms of the initial conditions $r_{0}$ and $p_{0}.$ It is, however, convenient to keep the expressions of $p_{i}$ in the intermediary step, in terms of $r_{n}$, $n>i$, which is to be replaced, at the end, by its final expression in terms of the initial conditions. From the Eq. (\ref{rnA}) we may write
\be \l{rnj}
\frac{r_{n}}{r_{j}}=\prod_{i=j}^{n}(1+\beta p_{i})=1+\beta\sum_{i_{1}=j}^{n}p_{i_{1}}{\Big\{}1+\beta\sum_{i_{2}=i_{1}+1}^{n}p_{i_{2}}{\Big\{}1+\dots+\sum_{i_{n}=i_{n-2}+1}^{n}p_{n-1}{\Big\}}\dots{\Big\}},
\ee
and so,  from the Eq. (\ref{vj})
\be
\l{vnj}
p_{j}=p_{0}-\frac{q}{r_{n}}\sum_{i=1}^{j}\frac{r_{n}}{r_{i}}=p_{0}-\gamma\sum_{i=1}^{i}{\Big\{}1+\beta\sum_{i_{1}=i}^{n}p_{i_{1}}{\Big\{}1+\beta\sum_{i_{2}=i_{1}+1}^{n}p_{i_{2}}{\Big\{}1+\dots+\sum_{i_{n-1}=i_{n-2}+1}^{n}p_{n-1}{\Big\}}\dots{\Big\}}{\Big\}},
\ee
with $\gamma=\frac{q}{r_{n}}$. The Eqs. (\ref{dti}) and (\ref{dv}) replace the differential equation of the continuous fields whereas the $(n+1)$-term finite series  (\ref{rn0}) and (\ref{vnj}) replace their respective continuous solutions. As $n$ usually is a huge integer the successive sums may become quite involved  and so it is convenient to adopt a systematic approach using 
\be
 {n\choose k}=\cases{0,&if $n<k$;\cr
\cr
_1,&if $k=0$;\cr
\cr
\frac{n!}{(n-k)!k!},&if $n\ge k\ge0$,\cr}
\ee
 with
\be
\l{eeq}
\sum_{i=0}^{n-1}{i\choose k}={n\choose k+1},
\ee
\be
\l{eeq1}
\sum_{i=1}^{n}\equiv\sum_{i=0}^{n-1}+\delta_{i}^{n}-\delta_{i}^{0},
\ee
\be
\l{eeq2}
\sum_{i=0}^{n-1}i{i\choose k}=(k+1){n\choose k+2}+k{n\choose k+1},
\ee
\be
\l{eeq3}
\sum_{i=0}^{n-1}i^k{i\choose k'}=\frac{(k+k')!}{k'!}{n\choose k+k'+1}+{\hbox{smaller order terms}},
\ee
and other results derived from these \cite{gr-qc/0111030} for a systematic writing of each term of this series in terms of combinatorials ${n\choose k}$. For $n>>k$ the Eqs. (\ref{eeq}) and (\ref{eeq3}), may be written respectively as
\be
\l{eeq'}
\sum_{i=0}^{n-1}\frac{i^{k}}{k!}\approx\frac{n^{k+1}}{(k+1)!},
\ee
\be
\l{eeq4}
\sum_{i=0}^{n-1}i^{k+k'}{k'!}=\frac{(k+k')!}{k'!}\sum_{i=0}^{n-1}\frac{i^{k+k'}}{(k+k')!}\approx \frac{n^{k+k'+1}}{(k+k'+1)!}.
\ee

Let us introduce the following expansions
\be
\l{vs}
p_{n}=\sum_{s=0}^{s=n}\beta^s p^{(s)}_{n},
\ee
\be
\l{xsA}
x_{n}\equiv\frac{r_{n}}{r_{0}}=\sum_{s=0}^{s=n}\beta^s X^{(s)}_{n},
\ee
and 
\be
\l{vnjs}
N_{j}=\sum_{s=0}^{s=n}\beta^s N^{(s)}_{j},
\ee
where $N_{j}$ represents $p_{j}$ expressed in terms of $r_{n}$ and not of $r_{j}$, $N_{n}=p_{n}$, $N^{(s)}_{n}=p_{n}^{(s)}$.
Then we have, from Eqs. (\ref{vnj},\ref{vs},\ref{vnjs}),  
\be
\l{Vjn0}
N_{j}^{(0)}=p_{0}-\gamma\sum_{i=1}^{j}1=p_{0}-\gamma{j\choose 1},
\ee 
and for $n>k\ge1$,
\be
\l{Vnjs}
N_{j}^{(k)}= -\gamma\sum_{i=1}^{j}\sum_{m_{1}=i}^{n}{\Big\{}N_{m_{1}}^{(k-1)}+\sum_{m_{2}=m_{1}+1}^{n}{\big\{}(N_{m_{1}}N_{m_{2}})^{(k-2)}+\dots+\sum_{m_{k-1}=m_{k-2}+1}^{n}(N_{m_{1}}N_{m_{2}}\dots N_{m_{k}})^{(0)}{\big\}}\dots{\Big\}},
\ee
where 
\be
\l{VV}
(N_{m_{1}}N_{m_{2}})^{(k)}\equiv \sum_{s=0}^{s=k}N_{m_{1}}^{(s)}N_{m_{2}}^{(k-s)},
\ee
\be
\l{VVV}
(N_{m_{1}}N_{m_{2}}N_{m_{3}})^{(k)}\equiv \sum_{s=0}^{s=k}N_{m_{1}}^{(s)}(N_{m_{2}}N_{m_{3}})^{(k-s)},
\ee
and so on. For example
\be
N_{j}^{(1)}\approx -\gamma\sum_{i=1}^{j}\sum_{m_{1}=i}^{n-1}N_{m_{1}}^{(0)}=-\gamma\sum_{i=1}^{j}\sum_{m_{1}=i}^{n}(p_{0}-\gamma{j\choose 1})
\approx-\gamma{\Big\{}p_{0}{\Big[}{n\choose 1}{j\choose 1}-{j\choose 2}{\Big]}-\gamma{\Big[}{n\choose 2}{j\choose 1}-{j\choose 3}{\Big]}{\Big\}},
\ee

\be
p_{n}^{(1)}
=-\gamma\sum_{i=1}^{j}\sum_{m_{1}=i}^{n-1}N_{m_{1}}^{(0)}=-\gamma\sum_{i=1}^{j}\sum_{m_{1}=i}^{n}(p_{0}-\gamma{j\choose 1})
\approx-\gamma n^2(\frac{p_{0}}{2}-\frac{n\gamma}{3}),
\ee
\be
N_{j}^{(2)}=-\gamma\sum_{i=1}^{j}\sum_{m_{1}=i}^{n-1}{\Big\{}N_{m_{1}}^{(1)}+\sum_{m_{2}=m_{1}+1}^{n}{\big\{}(N_{m_{1}}N_{m_{2}})^{(0)}{\big\}}{\Big\}}\approx-\gamma{\Big\{}p_{0}[{n\choose 1}{j\choose 1}-{j\choose 2}]-\gamma[{n\choose 2}{j\choose 1}-{j\choose 3}]{\Big\}},
\ee
\be
p_{n}^{(2)}=-\frac{\gamma n^3}{3}(\frac{p_{0}^2}{2}-\frac{5np_{0}\gamma}{4}+\frac{4n^2\gamma}{5})
\ee

$$N_{j}^{(3)}= -\gamma\sum_{i=1}^{j}\sum_{m_{1}=i}^{n-1}{\Big\{}N_{m_{1}}^{(2)}+\sum_{m_{2}=m_{1}+1}^{n-1}{\big\{}(N_{m_{1}}N_{m_{2}})^{(1)}+\sum_{m_{3}=m_{2}+1}^{n-1}(N_{m_{1}}N_{m_{2}}N_{m_{3}})^{(0)}{\big\}}{\big\}}{\Big\}},$$
and so on. 
Also, from Eqs. (\ref{rn0},\ref{xsA},\ref{vnjs}) we have
$$
X_{n}^{(k)}=\sum_{m_{1}=0}^{n-1}{\Big\{}N_{m_{1}}^{(k-1)}+\sum_{m_{2}=m_{1}+1}^{n-1}{\Big\{}(N_{m_{1}}N_{m_{2}})^{(k-2)}+\sum_{m_{3}=m_{2}+1}^{n-1}{\Big\{}(N_{m_{1}}N_{m_{2}}N_{m_{3}})^{(k-3)}+\dots 
$$
\be
\l{Xn}
\dots+\sum_{m_{k-1}=m_{k-2}+1}^{n-1}(N_{m_{1}}N_{m_{2}}\dots N_{m_{k}})^{0}{\Big\}}\dots{\Big\}}.
\ee
For example,

\be
\l{Xn0}
X_{n}^{(0)}=1,
\ee

\be
\l{Xn1}
X_{n}^{(1)}=\sum_{m_{1}=0}^{n-1}N_{m_{1}}^{(0)}=\sum_{m_{1}=0}^{n-1}(p_{0}-\gamma{m_{1}\choose 1})=p_{0}{n\choose 1}-\gamma{n\choose 2}\approx n(p_{0}-\frac{n\gamma}{2})=\frac{p_{0}^2-{p_{n}^{(0)}}^2}{2\gamma},
\ee
$$
X_{n}^{(2)}=\sum_{m_{1}=0}^{n-1}{\Big\{}N_{m_{1}}^{(1)}+\sum_{m_{2}}^{n-1}{\Big\{}(N_{m_{1}}N_{m_{2}})^{(0)}{\Big\}}{\Big\}}=p^{2}_{0}{n\choose 2}-\frac{p_{0}\gamma}{6}{\Big\{}5n^{3}-9n^{2}+4n{\Big\}}+\frac{\gamma^{2}_{n}}{6}{\Big\{}2n^{4}+5n^{3}+4n^{2}-n{\Big\}}\approx,
$$
\be
\approx \frac{n^2}{2}(p_{0}^2-\frac{5}{3}np_{0}\gamma+\frac{2}{3}n^2\gamma^2)=\frac{p_{n}^{(0)}p_{n}^{(1)}}{\gamma}
\ee

$$
X_{n}^{(3)}=\sum_{m_{1}=0}^{n-1}{\Big\{}N_{m_{1}}^{(2)}+\sum_{m_{2}=m_{1}+1}^{n-1}{\Big\{}(N_{m_{1}}N_{m_{2}})^{(1)}+{\Big\{}\sum_{m_{3}=m_{2}+1}^{n-1}(N_{m_{1}}N_{m_{2}}N_{m_{3}})^{(0)}{\Big\}}{\Big\}}{\Big\}},
$$ and so on.

Eqs. (\ref{Vnjs},\ref{Xn}) are recursion relations that allow the complete determination of both finite series (\ref{vs}) and (\ref{xsA}). So, $p_{n}$ and $r_{n}$ can be determined from the knowledge of $p_{0}$ and $r_{0}$. On the other hand, considering that generally $n$ is a very large number we have considered only the dominant contribution from each term in each series. Then we have
\be
\l{svn}
p_{n}\simeq{\Big \{}p_{0}-n\gamma{\Big \}}-\beta{\Big \{}n^2\gamma(\frac{p_{0}}{2}-\frac{n\gamma}{3}){\Big \}}-\beta^2{\Big \{}\frac{n^3\gamma}{3}(\frac{p_{0}^2}{2}-\frac{5}{4}p_{0}n\gamma+\frac{4}{5}n^2\gamma){\Big \}}-\dots
\ee
and
\be
\l{sxn}
x_{n}\simeq1+\beta{\Big \{}n(p_{0}-\frac{n\gamma}{2}){\Big \}}+\beta^2{\Big \{}\frac{n^2}{2}(p_{0}^2-\frac{5}{3}np_{0}\gamma+\frac{2n^2\gamma^2}{3}){\Big \}}+\dots
\ee
or, expressing $x_{n}$ in terms of $p_{n}^{(s)}$,
$$
\frac{r_{n}}{r_{0}}=1+\beta{\Big \{}\frac{p_{0}^2-p_{n}^{(0)^{2}}}{2\gamma}{\Big \}} -\beta^2{\Big \{}\frac{2p^{(0)}_{n}p_{n}^{(1)}}{2\gamma}{\Big \}}-\beta^3{\Big\{} 
\frac{p_{n}^{(1)^{2}}+2p_{n}^{(0)}p_{n}^{(2)}}{2\gamma}{\Big \}}-\dots=
$$
\be
\l{xsn1}
=1+\beta{\Big \{}\frac{p_{0}^2-(p_{n}^{(0)}+\beta p_{n}^{(1)}+\beta^2 p_{n}^{(2)}+\dots)^2}{\gamma}{\Big \}}=1+\frac{\beta}{\gamma}(\frac{p_{0}^2-p_{n}^{2}}{2}),
\ee
The Eq. (\ref{xsn1}) has been obtained with the neglecting of smaller contributions, the non-dominant terms in the ${n\choose k}$'s. The exact expression for $x_{n}$, including all contributions, can be written as 
\be
\l{KUn}
x_{n}=1+\frac{\beta}{\gamma}(\frac{p_{0}^2-p_{n}^{2}}{2})+\sum_{s=1}\beta^{s}{X'}_{n}^{(s)},
\ee
where ${X'}_{n}^{(s)}$ represents all the non-dominant contributions from $X_{n}^{(s)}$. Then Eq. (\ref{KUn}) can be rearranged as
\be
\l{Xlinha}
(\frac{p_{0}^2}{2m}-\frac{GMm}{r_{0}})-(\frac{p_{n}^2}{2m}-\frac{GMmr_{n}})=-\sum_{s=1}^{s=n}\frac{\beta^{s-1}}{m}\gamma{X'}_{n}^{(s)}.
\ee
We notice the appearing of the effective potential
$U(r)=-\frac{q}{m^2\beta}\frac{1}{r}$,
which is the expression of a conservative field at the measure that the non-dominant contributions ${X'}_{n}^{(s)}$ can be neglected. Energy is always conserved but this standard expression for the gravitational potential is exact only in the limit of $n\rightarrow\infty,$ which would be the only way of justifying, in absolute terms, the neglecting of the right-hand-side of Eq. (\ref{Xlinha}). So, $U(r)$ is just an idealized useful limiting concept. On the other hand, the RHS of Eq. (\ref{Xlinha}), for ${X'}_{n}^{(s)}>0$ corresponds to an extra attractive potential.
From Eq. (\ref{Xn1}) we see that $\frac{n\gamma}{2}$ is the neglected term in $X_{n}^{(1)}$. Therefore, we have
\be
{X'}_{n}^{(1)}=\frac{n\gamma}{2}=\frac{nq}{2r_{n}},
\ee
and so that the right-hand-side of Eq. (\ref{Xlinha}) can be written as an extra generalized potential
\be
\l{Xlinha1}
\delta U(r_{n})=-\sum_{s=1}^{s=n}\beta^{s-1}\frac{q}{mr_{n}}{X'}_{n}^{(s)}=-\frac{q^{2}n}{2mr_{n}^{2}}
+{\cal{O}}(\beta).
\ee
From Eq. (\ref{KUn}) we have 
\be
r_{n}\approx\frac{r_{0}}{1-\frac{\beta r_{0}}{2q}(p_{0}^2-p_{n}^2)}+\delta r_{n}, \quad\delta r_{n}=-\frac{nq\beta}{2},
\ee
from which we get
\be
\l{rnA1}
\frac{dr_{n}}{dn}\approx-\frac{\beta p_{n}r_{n}^2}{q}\frac{dp_{n}}{dn}.
\ee
For $n>>1$ the derivative is a good approximation so that we can write from Eq. (\ref{vj})
\be
dp_{n}=-\frac{q}{r_{n}}dn
\ee
and, therfore
\be
\frac{dr_{n}}{dn}=\beta p_{n}r_{n}.
\ee
On the other hand, from Eq. (\ref{Dt0}) we get 
\be
t_{n}= t_{0}+m\beta\sum^{n-1}_{i=0}\frac{1} r_{i},
\ee
and 
\be
dt_{n}\approx\beta mr_{n}dn.
\ee
Then
\be
\frac{dr_{n}}{dt_{n}}\approx\frac{p_{n}}{m}\qquad{\hbox{and}}\qquad\frac{d\delta r_{n}}{dt_{n}}\approx\frac{q}{2mr_{n}},
\ee
and 
\be
\frac{d^2r_{n}}{dt^2_{n}}\approx\frac{1}{m}\frac{dp_{n}}{dt_{n}}\approx\frac{1}{m}\frac{dp_{n}}{dn}\frac{dn}{dt_{n}}=-\frac{GM}{r_{n}^2}\qquad{\hbox{and}}\qquad\frac{d^2\delta r_{n}}{dt^2_{n}}\approx-\frac{qp_{n}}{2m^2r^2_{n}},
\ee
which shows that the continuous description is entirely generated by the discrete dominant contributions and that the sub-dominant ones generate new physics that may or may not be observed in nature. 
The extra interaction, in a first order approximation gives a positive increment 
$\delta r_{n}$ to the expected (from the continuous approximation) value of $r_{n}$. Whereas it is always attractive for the second alternative, for the first one, depending on the initial conditions ($p_{0}$ and $r_{0}$), the extra interaction may be repulsive: the actual gravitational attraction is weaker (alternatively, stronger) than its continuous description. The extra interaction has this clear physical meaning. It is a correction to the excess contained in approximating the interaction by the effective continuous potential. Only precise, carefully done experiments must give the final answer. In contradiction to all other fundamental interactions the great riddles associated to gravity are in the large distance limit where the field is extremely weak. They are not easily fitted with a continuous interaction without evoking new forms of interaction or of matter. In a natural way discrete interactions introduce modifications to the standard physics in the very large and in the very small distance limits, and it is only in these asymptotic limits that these modifications can be experimentally detected. Outside them $n$ is huge enough for validating the continuous approximation but not enough for turning detectable the subdominant terms ($\delta x_{n}^{(s)}$).

\end{document}